\documentclass[12pt]{article}
\begin{document}
\input epsf
\baselineskip=15pt

\newcommand{\be}{\begin{equation}}
\newcommand{\ee}{\end{equation}}
\newcommand{\bq}{\begin{eqnarray}}
\newcommand{\eq}{\end{eqnarray}}
\newcommand{\x}{{\bf x}}
\newcommand{\y}{{\bf y}}

\newcommand{\p}{\varphi}
\newcommand{\del}{\nabla}
\begin{titlepage}
\vskip1in
\begin{center}
{\large\bf VEV's and condensates from the Schr\"odinger Wave-functional.}
\end{center}
\vskip1in
\begin{center}
{\large David Nolland

Department of Mathematical Sciences

University of Liverpool

Liverpool, L69 3BX, UK}

\vskip.5in

e-mail: {\tt nolland@liverpool.ac.uk}

\end{center}
\vskip1in
\begin{abstract}
\noindent We show how VEV's and condensates can be read off from
the Schr\"odinger wave-functional without further calculation.
This allows us to study non-perturbative physics by solving the
Schr\"odinger equation. To illustrate the method we calculate
fermion condensates from the exact solution of the Schwinger
model, and other (1+1) dimensional models. The chiral condensate
is seen to be a large-distance effect due to propagators
reflecting off the space-time boundary.
\end{abstract}

\end{titlepage}


\section{Introduction}
The non-perturbative calculation of expectation values is of
central importance to QFT, as we learn from the study of
everything from quark confinement and dynamical symmetry breaking
to string/gauge theory duality. Instanton approximations, lattice
gauge theory, and the study of supersymmetry all give us important
insights into non-perturbative physics, but as yet, methods for
continuum calculations in realistic models are extremely limited.

Another potential source of techniques for QFT is the functional
Schr\"odinger equation. The wave-functional solutions of this
equation give us a new perspective on insights generated by other
methods, and provide some new calculational tools.

We still need to know how to read off interesting physics from the
wave-functional. The calculation of expectation values from the
wave-functional is something of a vexed issue, since taking the
inner-product of an operator with respect to a QFT state formally
involves functional integrations, and in a non-Gaussian theory,
this requires a diagrammatic expansion and complicated
calculations, even if the wave-functional is already known.

It turns out that this is misleading. In this paper, we will show
how expectation values can be read off directly from the
wave-functional, without further integrations. We will show how
chiral symmetry breaking appears in this context in
simple models.

In a previous paper \cite{d2}, we solved the Schr\"odinger
equation for the Schwinger model at zero and finite temperature.
We will use this to illustrate our prescription for obtaining
expectation values. Because the theory is non-Gaussian, exhibits
many interesting phenomena such as a chiral condensate and phase
transition, but remains exactly solvable, it provides a good
illustration of general principles.

The non-zero value of the chiral condensate is a large-distance
effect: the wave-functional is written in terms of correlation
functions on a space-time boundary, while VEV's correspond to
correlation functions in the bulk. As the boundary is taken to
infinity, propagators reflecting off it can still give a non-zero
correction to the perturbative VEV's. This correction is given by
a large-distance limit of the boundary correlation functions.

This gives a mechanism whereby large-distance effects can be
responsible for the dynamical generation of condensates. It would
be interesting to see if other examples of dynamical symmetry
breaking, such as confinement, could be understood in this way.

 \vfill\eject

 \section{The Schr\"odinger wave-functional}

The primary object of study in our investigations is the
Schr\"odinger wave-functional, which can be interpreted as the
density matrix between energy eigenstates at a finite temperature
$T=1/k\tau$. If we let $T\rightarrow0$ then we get the vacuum
wave-functional. We will represent the Schr\"odinger functional as a
matrix element of the operator $e^{-\hat H\tau}$:
 \be
 \Psi[\psi_1,\psi_2]=\langle\psi_1|e^{-\hat H\tau}|\psi_2\rangle.
 \ee
This satisfies the time dependent Schr\"odinger equation $\hat
H\Psi=-\tau\Psi$, and can be interpreted as the path integral of
the exponentiated action over a Euclidean space-time region
bounded by surfaces at $t=0$ and $t=\tau$, on which the boundary
conditions are specified by the arguments $\psi_1$ and $\psi_2$
respectively.

Imposing the boundary conditions corresponds to introducing
sources on the boundary which diagonalize a maximal set of
canonical field variables. The conjugate variables are represented
by functional differentiation, and it is in such a representation that
concrete calculations will be performed.

For bosons we can select Dirichlet or Neumann conditions, or in
general some combination thereof. For fermions we found in
\cite{d2} a set of local projection operators that can be used to
specify the diagonal components. These boundary
conditions have several advantages, for example:

\begin{enumerate}
\item{They guarantee that Gauss' law is obeyed by the wave-functional.}
\item{When they are imposed the Dirac operator has no zero modes, which simplifies the
quantization considerably.}
\item{They provide a set of equivalent representations for the
wave-functional which may be used interchangeably.}
\end{enumerate}

Expectation values are obtained in the following manner: All
operators have a well-defined action on the wave-functional, and
their expectation value is the functional trace of the resulting
object. From the path-integral point of view this corresponds to
operator insertions on the boundary, followed by a functional
integration that glues the two boundaries together. This is
equivalent to the usual prescription which inserts the operators
on a cylinder of radius $\tau$. If we started with the vacuum
functional, the procedure is much the same; we glue together two
copies of the half-plane with suitable boundary insertions, and
integrate over boundary values. The boundaries disappear, and full
Euclidean invariance is restored.

Up to certain simple rescalings the wave-functional can itself be
interpreted as a generating functional of expectation values or
equal-time correlation functions. Then the (usually non-Gaussian)
functional integrations described in the last paragraph are
unnecessary, and most of the physically interesting information
can be read off from the wave-functional without too much further
work.

For simplicity we will consider in this paper mostly cases where
the renormalization of the Schr\"odinger representation requires
no additional counterterms on the boundary, apart from the usual
ones. In gauge theories (with or without fermions) this is
guaranteed by gauge invariance.

We will need to take account of boundary effects, ie. relate
correlation functions on the boundary to those deep in the bulk.
This will enable us to account for large-distance effects which
are responsible for the dynamical generation of condensates that
are responsible for chiral symmetry breaking. We speculate that a
similar phenomenon might be involved in the dynamical generation
of condensates leading to confinement.

\section{Calculation of correlation functions}

The vacuum functional can be written in the form:

\be
\Psi[\p]=exp(-W[\p])=\int D\phi\exp(-S+S_B)
\ee
where $S$ is the euclidean action, and $S_B$ is a boundary term imposing the
appropriate boundary conditions. These boundary conditions can be changed by
performing a functional Fourier transform---for example, if we started with the
"field" representation where $\phi$ is diagonal, then

\be
\tilde\Psi[\tilde\p]=\int D\p\Psi[\p]e^{i\p\tilde\p}
\ee
is the corresponding vacuum functional in the "momentum" representation where
the momenta conjugate to $\phi$ are diagonal.

Now according to the prescription for obtaining VEV's, we can write
\be
\exp(-F[J])=\int D\p\ |\Psi[\p]|^2\ e^{iJ\p},
\ee
where $F[J]$ is the generating functional for connected equal-time
correlation functions, so that the vacuum expectation of some function $G[\phi]$ is
given by
\be
G[{\delta\over\delta J}]\exp(-F[J])|_{J=0}.
\ee

Now for a wide range of interesting cases, the wave functional can be shown to be real,
in which case
\be
|\Psi[\p]|^2=\exp(-2W[\p])
\ee
Compare the two objects:
\bq
\exp(-F[J])&=&\int D\p \exp(-2W[\p])e^{iJ\p}\nonumber\\
\exp(-\tilde W[\tilde\p])&=&\int D\p\exp(-W[\p])e^{i\tilde\p\p}
\eq
If we can find a scaling of fields, coupling constants and other parameters that effects the transformation
$W[\p]\rightarrow2W[\p]$, then we can essentially identify the wave-functional in the second line
with the generating functional in the first.

Let us see how this works in practice. For a free theory, $W[\p]$
is quadratic, and (up to an overall constant) we can achieve the
desired effect by rescaling the argument: $\p\rightarrow\sqrt2\p$.
Then \be F[J]=\tilde W[J/\sqrt2] \ee More generally, this
rescaling will give the correct quadratic term for {\it any}
theory, so that the correct two-point correlators can easily be
extracted from the appropriate vacuum functional. This technique
is similar to the "plasma analogy" that is of importance in the
study of the fractional quantum Hall effect \cite{fref} and other
phenomena in statistical physics.

We can generalize this to non-free theories as follows. Consider a
theory with a mass gap. For slowly varying sources, the vacuum can
be expanded in integrals over space of local functions of the
fields and their derivatives at a single space-time
point.\footnote{This is a simplification. It may be necessary to
rewrite the vacuum functional in terms of other variables
\cite{nair}, or even to express it as a limit of some regulated
object \cite{d2} in order to achieve this result. But some sort of
local expansion can generally be found.}

Since $W[\p]$ is dimensionless, the integrand has dimension
$(length)^{-d}$, where $d$ is the dimension of space. We can
rescale all fields and dimensionful parameters according to their
dimension in such a way that $W[\p]\rightarrow2W[\p]$. Note that
we will also require derivatives to scale, which means scaling the
arguments of the fields in an appropriate way. The required
prescription is to ensure that objects of dimension
$(length)^\alpha$ scale as $2^{-\alpha/d}$. These scalings
translate into a well-defined scaling prescription for $\tilde
W[J]$ that makes it equal to $F[J]$.

The existence of a local expansion is not really necessary for
this procedure to work; it is enough that the vacuum functional is
derived from a local action. The point is that an appropriate
rescaling of dimensionful parameters is enough to account for the
difference between the vacuum functional and the generating
functional of equal time correlation functions.

\section{Large distance corrections to bulk correlation functions}

To illustrate how large distance effects can modify correlation
functions in the bulk, consider a particle propagating between two
points A and B in the interior of some bounded Euclidean region
$\tau\le t\le\infty$. The propagator contains contributions from
the straight path from A to B, as well as the path that reflects
off the boundary at $t=\tau$.

Now suppose that we restore Euclidean invariance by taking $\tau
\rightarrow-\infty$. Usually the latter contribution to the
propagator is suppressed, eg. for a free particle, where the
action is proportional to the path length. An exception to this is
the chiral condensate, which is non-zero only for an odd number of
reflections.

How does this affect the calculation of expectation values?
Suppose we have a theory containing complex fermions $\psi$ and
$\psi^\dagger$. On the boundary $t=\tau$ we will impose the
boundary conditions discussed in \cite{d2}, diagonalising
$Q_+\psi$ and $\psi^\dagger Q_-$, where $Q_\pm={1\over2}(1\pm Q)$
where $Q=a\gamma^0\pm b\gamma^1+c\gamma^5$ and $a^2+b^2+c^2=1$.

Consider the chiral condensate $<\bar\psi(x,0)\psi(x,0)>$.
We can rewrite this as \be <\bar\psi(x,0)
Q_+\psi(x,0)>+<\bar\psi(x,0)
Q_-\psi(x,0)>.\label{cc}\ee These expressions are to be
interpreted as propagators between coinciding points in the
interior of the bounded Euclidean region, and as
$\tau\rightarrow\infty$ they should tend to the vacuum expectation
values of the corresponding fermion currents.

Now consider the first term in (\ref{cc}). In the cases we will
consider, the dominant contribution to this term comes from the
shortest path via the boundary at $t=\tau$. We will choose
$Q=\gamma^5$, so that $\bar\psi Q_+=-\psi^\dagger Q_-\gamma^0$.
According to the method of images, and since $\psi^\dagger Q_-$
and $Q_+\psi$ are freely integrated over on the boundary, we have
\be <\bar\psi(x,0) Q_+\psi(x,0)>=<\psi^\dagger(x+2\tau,\tau)
Q_-\gamma^0Q_+\psi(x,\tau)>. \label{cc2}\ee The second term in
(\ref{cc}) is identical, but with $Q=-\gamma^5$, so that
$Q_\pm\rightarrow Q_\mp$. As we take the limit
$\tau\rightarrow\infty$, we see that the chiral condensate is
given by the large-distance limit of the boundary propagators.

Having re-expressed the chiral condensate in terms of boundary
correlation functions of non-diagonal fields, we can now use the
results of the last section to extract the value of the chiral
condensate from the vacuum functional. In \cite{d2} we found the
vacuum functional for the Schwinger model at zero and finite
temperature. Here we quote the result, generalising it for
amusement to the case of multiple fermion flavours.

The representation is:\footnote{The source for the electromagnetic
field can be gauged away completely, and the vacuum is expressed
as a function of fermion fields only.}

\be Q_+\hat\psi\sim u,\qquad \hat\psi^\dagger Q_-\sim \tilde u
\ee

\be Q_-\hat\psi\sim {\delta\over\delta \tilde u},\qquad
\hat\psi^\dagger Q_+\sim {\delta\over\delta u} \ee

and the vacuum wave-functional is

\bq \Psi[u,\tilde u]&=&\sum_{a=0}^\infty{1\over
a!}\prod_{n=0}^a\Biggl[{2\over\pi}\int dx_n dy_n \tilde
u(x_n)\gamma^0 u(y_n){\cal P}{1\over
x_n-y_n}\nonumber\\
&&\times\exp\left\{\sum_{i,j=1}^a\Phi(x_i-y_i)-\sum_{j>i=1}^a[\Phi(x_i-x_j)+\Phi(y_i-y_j)\right\}\Biggr],
\label{vac}\eq

where

\be \Phi(x)={1\over
N_f}\int{dp\over2\pi}\left({1\over|p|}-{\sqrt{p^2+m^2}\over
p^2}\right)(1-\cos(px)) \ee

for the Schwinger model at zero temperature, and

\be \Phi(x)={1\over N_f}\int{dp\over2\pi}\left(\coth(p\tau){1\over
p}-\coth(\sqrt{p^2+m^2}\tau){\sqrt{p^2+m^2}\over
p^2}\right)(1-\cos(px)) \ee

for the Schwinger model at temperature $T=1/k\tau$, where
$m=\sqrt{N_fe^2\over\pi}$ is the Schwinger mass for $N_f$
flavours.

According to our scaling prescription, we can calculate the
boundary propagator in (\ref{cc2}) as

\bq <\psi^\dagger Q_-\gamma^0
Q_+\psi>&=&\lim_{x-y\rightarrow\infty}{\delta\over\delta
u(x)}\gamma^0{\delta\over\delta \tilde
u(y)}\Psi[{1\over\sqrt2}u,{1\over\sqrt2}\tilde u]|_{u,\tilde
u=0}\nonumber\\&=& N_f{\rm
tr}(Q_-\gamma^0Q_+\gamma^0)\lim_{x\rightarrow\infty}{1\over \pi
x}\exp(\Phi(x)).\eq

$\Phi(x)$ has the asymptotic form

\be \Phi(x)\sim {1\over N_f}(\gamma+\ln(mx/2))+O(e^{-mx}) \ee

for both zero and finite temperature, and
tr$(Q_\mp\gamma^0Q_\pm\gamma^0)=1/2$. Putting this all together,
we have

\be <\bar\psi\psi>={me^\gamma\over2\pi} \ee

for $N_f=1$, and $<\bar\psi\psi>=0$ for $N_f>1$.

Because of the anomaly, the vacuum angle can be altered by a
chiral rotation of the fermion fields:

\be \hat\psi\rightarrow e^{i\theta\gamma^5/2}\hat\psi,\qquad
\hat\psi^\dagger\rightarrow\hat\psi^\dagger e^{-i\theta\gamma^5/2}
\ee

Under this transformation we have for $N_f=1$

\bq <\bar\psi\psi>&\rightarrow&<\bar\psi
e^{i\theta\gamma^5}\psi>\nonumber\\ &=&e^{i\theta}<\bar\psi
Q_+\psi>+e^{-i\theta}<\bar\psi Q_-\psi>\nonumber\\
&=&{me^\gamma\over2\pi}\cos\theta \eq

The Schwinger model in a finite box exhibits a second order phase
transition at $T=0$, which has been studied numerically
\cite{durr}. The above analysis is readily extended to this case.

Consider a more general model with a four-fermi interaction

\be {1\over2}\int dx\int dy
\psi^\dagger(x)\psi(x)u(x-y)\psi^\dagger(y)\psi(y) \ee

The Schwinger model corresponds to $u(x)=-e^2|x-y|$, and in
general the vacuum wave-functional is given by (\ref{vac}) with

\be \Phi(x)=\int{dp\over2\pi}{1\over|p|}\left(1-{\tilde
u(p)\over\pi} \right)^{1\over2}(1-\cos(px)) \ee

with $\tilde u(p)$ the Fourier transform of $u(x)$. To get a
non-zero chiral condensate, the large-distance behaviour of
$\Phi(x)$ must have the same leading-order logarithmic behaviour
as for the Schwinger model. The massive Schwinger model is an
example. Note that the actual value of the condensate is
determined by the small-distance behaviour of $\Phi(x)$ and may in
fact be extracted from the vacuum energy using mass perturbation
theory \cite{adam}.

\section{Discussion}

The disadvantage of our prescription for obtaining VEV's is that
it only works for fields that are non-diagonal in our chosen
representation for the vacuum functional. Thus we cannot
straightforwardly obtain VEV's that are mixed functions of fields
and their canonical conjugates, such as the gluon condensate.
Also, we are in general tied to a specific representation.

For example, the vacuum wave-functional for (2+1) dimensional
Yang-Mills is given in \cite{nair} for the A-representation
($A_\mu$ diagonal), but the corresponding calculation in the
electric representation is less straightforward (though two-point
correlators are easily obtained by restricting to the abelian part
of the wave-functional.

Nevertheless, it would in principle be possible to calculate
expectation values of physically relevant operators such as the
Wilson loop from the appropriate wave-functional, if it could be
found. It is likely that large-distance effects, such as that
responsible for chiral condensates in the simple models we have
considered, would need to be taken into account in such
calculations.

\end{document}